\documentclass[twocolumn,prd,showpacs,preprintnumbers,superscriptaddress]{revtex4} 
\usepackage{graphicx} 
 
\begin{document} 
\newcommand{\AWEC}{AWEC} 
\newcommand{\WEC}{WEC} 
\newcommand{\QI}{QI} 
\newcommand{\n}{$(-)$} 
\newcommand{\p}{$(+)$} 
\newcommand{\T}[2]{\ifmmode T_{#1#2}\else $T_{#1#2}$\fi} 
 
\preprint{gr-qc/0109061} 
 
\title{Constraints on spatial distributions of negative energy} 
 
\author{Arvind Borde} 
 \email[Email: ]{arvind.borde@liu.edu} 
 \affiliation{Institute of Cosmology, Department of Physics and Astronomy\\ 
         Tufts University, Medford, MA~02155, USA.} 
 \affiliation{Theoretical and Computational Studies Group, Technology Center\\ 
              Southampton College, NY~11968, USA.} 
\author{L.H. Ford} 
 \email[Email: ]{ford@cosmos.phy.tufts.edu} 
 \affiliation{Institute of Cosmology, Department of Physics and Astronomy\\ 
         Tufts University, Medford, MA~02155, USA.} 
\author{Thomas A. Roman} 
 \email[Email: ]{roman@ccsu.edu} 
 \altaffiliation[\protect\\ Permanent address: ]{Department of Physics and Earth Sciences, 
         Central Connecticut State University, New Britain, CT~06050, USA.} 
 \affiliation{Institute of Cosmology, Department of Physics and Astronomy\\ 
         Tufts University, Medford, MA~02155, USA.} 
 
\date{ December  17, 2001} 
 
\begin{abstract} 
This paper initiates a program which seeks to study the allowed spatial 
distributions of negative energy density in quantum field theory. Here 
we deal with free fields in Minkowski spacetime. Known restrictions on time 
integrals of the energy density along geodesics, the averaged weak energy 
condition and quantum inequalities are reviewed. These restrictions are 
then used to discuss some possible constraints on the allowable spatial 
distributions of negative energy. We show how some geometric configurations 
can either be ruled out or else constrained. We also construct some explicit 
examples of allowed distributions. Several issues related to the allowable 
spatial distributions are also discussed. These include spacetime averaged 
quantum inequalities in two-dimensional spacetime, the failure of 
generalizations of the averaged weak energy condition to piecewise geodesics, 
and the issue of when the local energy density is negative in the frame of 
all observers. 
\end{abstract} 
 
 
\maketitle 
 
\section{\label{sec:intro}Introduction} 
 
The energy density of all observed forms of classical matter is non-negative. 
However, quantum field theory has the remarkable property that the local 
energy density can be negative. This violates the weak energy condition 
(WEC) which postulates that the local energy density is non-negative for 
all observers. Its formal statement is that a stress tensor $T_{\mu\nu}$ 
satisfies the WEC provided that 
\begin{equation} 
T_{\mu\nu}\, u^\mu\, u^\nu \, \geq 0 \, ,  \label{eq:wec} 
\end{equation} 
for all timelike vectors $u^\mu$. 
 Negative energy densities raise the possibility of a 
variety of exotic phenomena, including violations of the second law of 
thermodynamics \cite{F78,D82}, traversable wormholes \cite{MT88,MTY88} 
and ``faster-than-light'' travel \cite{A94,O98,GW}, creation of naked 
singularities \cite{FR90,FR92}, and avoidance of singularities in gravitational 
collapse. However, in the case of inflationary cosmology it has recently been 
found that violations of the WEC do not allow one to avoid initial singularities 
\cite{BGV}.

\subsection{\label{subsec:intro:review}Brief review of averaged weak energy 
                           conditions and quantum inequalities} 
 
The interest attached to the effects of negative energy has stimulated 
the study of constraints on the magnitude and extent of WEC violations. 
Although the energy density at a point can be arbitrarily negative, 
there are several integral constraints which we will briefly review
for the case of flat spacetime. One constraint is that the 
integral of the energy density over all space, the Hamiltonian, is 
bounded below. Other constraints involve integrals over the world line of 
an observer, a construction first introduced by Tipler~\cite{T78}. 
One such constraint is the averaged weak energy condition 
(AWEC), which states that the integral of the energy density seen by a 
geodesic observer is non-negative: 
\begin{equation} 
\int_{-\infty}^{\infty} T_{\mu\nu}\, u^\mu\, u^\nu \, d\tau \geq 0 \,. 
                                                  \label{eq:awec} 
\end{equation} 
Here $u^\mu$ and  $\tau$ are the observer's four-velocity and proper time, 
respectively. 
This condition has been proven to hold for a variety of free quantum 
field theories in boundary-free Minkowski spacetime. It does not always hold 
inside of a cavity in flat spacetime because the Casimir energy density 
can be negative. However, even in this case, observers at rest with 
respect to the cavity walls will see a modified version of Eq.~(\ref{eq:awec}) 
satisfied. This is the ``difference AWEC'' in which $T_{\mu\nu}$ is replaced by the 
difference between the expectation value of the stress tensor in an arbitrary 
quantum state and that in the vacuum state \cite{FR95,Mitch98}. The physical content 
of this statement is that although the local energy density can be made 
more negative than in the vacuum, the time-integrated energy density cannot. 
A related averaged energy condition is the averaged null energy 
condition (ANEC), in which the integration is along a null geodesic. It 
also holds in boundary-free Minkowski spacetime \cite{WY91}. The extent 
to which the AWEC and ANEC hold for quantum field theories in curved spacetime 
is less clear \cite{Y90,V95,FW96}. 
 
Although the AWEC imposes a significant constraint on negative energy, 
even stronger constraints are available in the form of ``quantum inequalities'', 
(QIs) \cite{F78,F91,FR95,FR97,FLAN,PF971,FE}. 
 These are lower bounds on time integrals of the energy density 
multiplied by a sampling function, $g(t)$: 
\begin{equation} 
\int_{-\infty}^{\infty} T_{\mu\nu}\, u^\mu\, u^\nu \,g(t) \, dt 
      \geq {\hat \rho}_{min}\,. 
\end{equation} 
The lower bound, ${\hat \rho}_{min}$, depends upon the sampling function 
and upon the spacetime. For the massless scalar field in two-dimensional 
Minkowski spacetime, Flanagan \cite{FLAN} has found the optimal bound: 
\begin{equation} 
 {\hat \rho}_{min} =  - \frac{1}{24 \,\pi} \, \int_{-\infty}^{\infty} \, 
\frac{{g'(t)}^2}{g(t)} \, dt  \,.     \label{eq:flanqi} 
\end{equation} 
 For the massless scalar field in four-dimensional Minkowski spacetime, 
Fewster and Eveson \cite{FE} have given a similar (but not necessarily 
optimal) bound: 
\begin{equation} 
 {\hat \rho}_{min} = -\frac{1}{16 \,{\pi}^2}\, 
\int_{-\infty}^{\infty} \,{\bigl( {g^{1/2}}''(t) \bigr)}^2 \,. 
\label{eq:feqi} 
\end{equation} 
 
If the sampling function has a characteristic width $t_0$, then the 
lower bounds are of the form 
\begin{equation} 
{\hat \rho}_{min} = \frac{C}{t_0^D} \,, 
\end{equation} 
where $D$ is the dimensionality of spacetime. The AWEC can be derived from 
the QIs as the limit in which $t_0 \rightarrow \infty$. 
The essential physical content of the QIs is that the larger a pulse of 
negative \n\ energy is, the closer in time it must be to a compensating pulse 
of positive \p\ energy. Consider for example $\delta$-function pulses 
of \p\ and \n\ energy where the energy density in an observer's frame 
is given by 
\begin{equation} 
\rho(t) = B \, [-\delta (t) + (1+\epsilon) \, \delta (t-T)] \,. 
\label{eq:pprofile} 
\end{equation} 
Here $T$ is the temporal separation of the pulses and $B$ is either 
the magnitude $|\Delta E|$ of the \n\ pulse in two dimensions, or the 
magnitude of its energy per unit area $|\Delta E|/A$ in four dimensions. 
It may be shown from the QIs \cite{FR99} that 
\begin{equation} 
T  \leq \frac{K_2}{|\Delta E| } \,, 
\label{eq:Tmax2d} 
\end{equation} 
in two dimensions and 
\begin{equation} 
T  \leq  K_4\,{\biggl(\frac{A}{|\Delta E| } \biggr)}^{1/3} \,, 
\label{eq:Tmax4d} 
\end{equation} 
in four dimensions, where the dimensionless constants $K_2$ and $K_4$ 
are typically less than unity. Thus as the strength $|\Delta E|$ of 
the \n\ pulse increases, its separation in time from the compensating 
\p\ pulse must decrease as an inverse power of  $|\Delta E|$. 
 
In Ref.~\cite{FR99} it is further shown that the QIs imply the phenomenon 
of {\it quantum interest}. As the temporal separation of the \n\ and \p\ 
pulses increases, within the limits set by Eqs.~(\ref{eq:Tmax2d}) and 
(\ref{eq:Tmax4d}), the degree of overcompensation must increase. Thus the 
parameter $\epsilon$ must be a monotonically increasing function of $T$. 
A  discussion of quantum interest for more general pulses was given by 
Pretorius \cite{P00} and by Fewster and Teo \cite{FT00}. 
 
\subsection{\label{subsec:intro:program}The program} 
 
The purpose of this paper is to initiate an exploration of the limits on 
the spatial distribution 
of \n\ energy. The worldline QIs summarized in the previous subsection 
provide one tool for this investigation. Clearly, for massless fields, 
the temporal separation 
of a pair of pulses as seen by an inertial observer is also a measure 
of their spatial separation. However, a more detailed picture is desirable. 
Several approaches can be pursued in the search for a description of 
the allowed spatial \n\ energy distributions. One is to seek generalizations 
of the QIs which involve averaging over space as well as time. Another 
approach is to use the AWEC and QIs to place constraints upon the allowed 
spacetime distributions. The approach seeks to extract as much information 
as possible from the requirement that the AWEC and QIs be satisfied along 
all timelike geodesics. A third approach is to examine the nature of 
distributions which are definitely allowed and can be explicitly 
constructed. All three approaches will be illustrated in this paper. 
 
\subsection{\label{subsec:intro:outline}Outline of this paper} 
This paper will deal entirely with \n\ energy distributions in flat 
spacetime. 
In Sec.~\ref{subsec:spatial:disturb}, we review and discuss two results 
which suggest that arbitrarily large amounts of \n\ energy can be concentrated 
in a given region of space. As a counterpoint, we show in 
Sec.~\ref{subsec:spatial:avgeqi} that in two spacetime dimensions, there 
are both spatial and spacetime averaged versions of the quantum inequalities. 
We next turn in Sec.~\ref{sec:fcdistr} to a discussion of several model 
\n\ energy distributions which the AWEC and QIs either forbid 
(Sec.~\ref{subsec:fcdistr:f}), or else quantitatively constrain 
(Sec.~\ref{subsec:fcdistr:c}). Section~\ref{sec:explicit} is devoted 
to the explicit construction of some informative examples of allowed 
distributions. In particular, the energy distribution of a massive scalar 
field in a single wavepacket mode squeezed state is used to illustrate the 
convoluted way in which \n\ and \p\ energy can be entwined. As part of this 
discussion, it is useful to distinguish between WEC violations in which 
the local energy density is negative for all observers (``strong'' violations), 
and those in which its sign depends upon the observer (``weak'' violations). 
The technical details of this distinction are elaborated in the Appendix. 
Section~\ref{sec:awec} explores the limits of the AWEC, and shows that 
it would not hold if one were to integrate along a piecewise geodesic 
path. Similarly, the ``difference AWEC'' need not hold for the quantum field 
stress tensor in a cavity in the case of an observer who passes through 
the cavity. This section also uses the cavity example to illustrate 
strong and weak WEC violations. Finally, our results are summarized and 
discussed in Sec.~\ref{sec:summary}. Units in which $\hbar = c = 1$ and 
a spacelike metric convention are used in this paper. 
 
\section{\label{sec:spatial}Difficulties with spatial bounds?} 
 
\subsection{\label{subsec:spatial:disturb}Two Disturbing Results} 
 
There are two disturbing results which might be construed as casting doubt 
upon the existence of constraints on the spatial distribution of negative 
energy. The first is an unpublished result of Garfinkle \cite{GAR}, 
who showed that 
the total energy contained within an imaginary box in Minkowski spacetime 
is unbounded below. Let us first give a more precise statement of this result. 
Consider a quantum field $\varphi$ in boundary-free Minkowski spacetime. 
By boundary-free, we mean that there are no physical boundaries upon which 
$\varphi$ must satisfy boundary conditions. Let $\rho = :T_{tt}:$ be the 
normal-ordered energy density operator for $\varphi$ on a $t = $ constant 
hypersurface. Now consider a volume $V$, e.g., the interior of an arbitrary 
rectangular box, and let 
\begin{equation} 
E = \int_V \langle \psi| \rho |\psi \rangle  \, d^3x 
\end{equation} 
be the energy inside this box in quantum state $|\psi \rangle$. 
This box is ``imaginary'' in the sense that there are no physical boundaries 
at the walls of the box. The Garfinkle result is that $E$ is unbounded 
below. That is, there exist states $|\psi \rangle$ for which $E$ is 
arbitrarily negative. Note that this would not happen if the box were 
a physical box on whose walls $\varphi$ must satisfy Dirichlet or Neumann 
boundary conditions. In this case, $E$ is the Hamiltonian for $\varphi$ 
within this cavity, and is bounded below by the Casimir energy of the 
cavity. 
 
\begin{figure} 
\includegraphics[scale=.35]{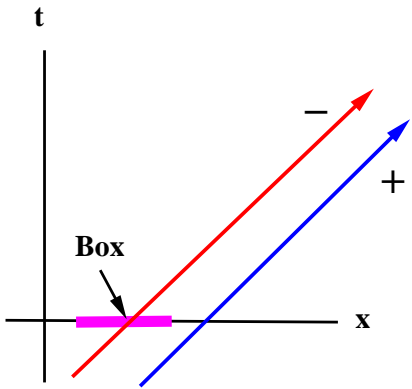} 
\caption{\label{fig:gar_box} The Garfinkle box is illustrated. A $\delta$-function pulse 
of $(+)$ energy 
has already passed through the box before time $t = 0$. At  time $t = 0$, 
a $\delta$-function pulse of $(-)$ energy is inside the box. 
The magnitude of this $(-)$ 
pulse is inversely related to the distance to the $(+)$ pulse at a fixed time. 
However, we can always arrange for the $(+)$ pulse to be just outside the box, 
and for the $(-)$ pulse to be just inside. Thus there can be an arbitrary 
amount of $(-)$ energy inside the box at $t = 0$.} 
\end{figure} 
 
\begin{figure} 
\includegraphics[scale=.35]{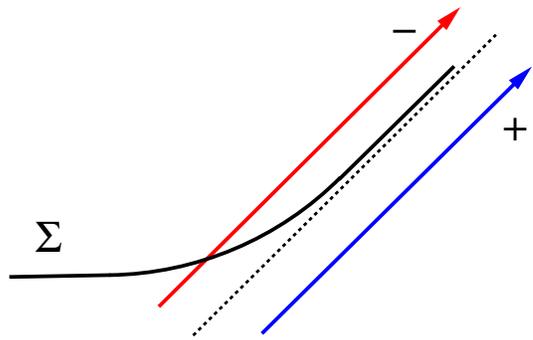} 
\caption{\label{fig:helf_slice} Here $\Sigma$ is a spacelike hypersurface which 
is asymptotic to a 
null surface, the dashed line. This allows $\Sigma$ to catch a 
$\delta$-function pulse of $(-)$ energy, while also avoiding the compensating 
$\delta$-function pulse of $(+)$ energy. In this way, the integrated energy 
over $\Sigma$ may be made arbitrarily negative.} 
\end{figure} 
 
A second disturbing result was given by Helfer \cite{Helfer96}, who showed 
that the integral of the energy density over a spacelike hypersurface 
can be unbounded below. Although this result applies to curved, as well as 
flat spacetime, let us focus on the case of Minkowski spacetime. Let 
$\xi^\mu$ be a timelike vector field on  Minkowski spacetime, and let 
$\Sigma$ be a spacelike hypersurface to which $\xi^\mu$ is orthogonal. 
Further let 
\begin{equation} 
\hat{H}(\xi,\Sigma) = \int \langle \psi| T^{\mu\nu}\, \xi_\mu \, \xi_\nu 
                      |\psi \rangle \, f \, dv \,. 
\end{equation} 
Here $dv$ is the volume element in $\Sigma$, and $f$ is a test function 
with compact support. The quantity $\hat{H}(\xi,\Sigma)$ is 
an energy operator (generalized  Hamiltonian) obtained by integrating 
the energy density in state $|\psi \rangle$ (times the test function $f$) 
over $\Sigma$. Helfer has shown that in general, $\hat{H}(\xi,\Sigma)$ 
is unbounded below. Note that in the limit that $f \rightarrow 1$ everywhere 
and $\xi^\mu$ is the timelike Killing vector, $\hat{H}(\xi,\Sigma)$ 
becomes the usual Hamiltonian, which has the lower bound of zero, attained 
in the Minkowski vacuum state. Note also that the Helfer result includes 
the Garfinkle result as the special case in which $\Sigma$ is 
a constant $t$ (Minkowski time) surface and $f$ approaches a step function 
which is $1$ inside the box and $0$ outside of it.

 Both of these results might lead one to conclude that there can be no 
bounds on the spatial distribution of negative energy which would be 
analogous to the temporal bounds given by the quantum inequalities. 
Thus it is desirable to understand the physical basis of  these 
results in more detail. 
 
Consider first the Garfinkle box. We can understand the unboundedness of 
the total energy $E$ in this box as arising from two factors: (1) The 
energy is measured at a precise instant in time, and (2) the walls of 
the box are sharply defined. This allows an arbitrary amount of negative energy 
to have entered the box by time $t$, while at this time excluding an even 
larger amount of positive energy which may be just outside of the box at 
time $t$. This situation is illustrated in  Fig.~\ref{fig:gar_box}.

The unboundedness of Helfer's $\hat{H}$ is harder to understand, although 
in particular cases one can give intuitive explanations similar to that 
in the  Garfinkle box case. Let the hypersurface $\Sigma$ be 
asymptotically null, as illustrated in Fig.~\ref{fig:helf_slice}. 
In this case, it is possible for the integral 
of the energy density over $\Sigma$ to include the contribution of an 
arbitrarily large negative energy pulse, but to omit that of an even larger 
positive pulse which preceded the negative pulse. Once we include the effects 
of the test function $f$, it is not necessary that $\Sigma$ be 
asymptotically null; it can level out and approach a  constant $t$ surface 
outside of the domain of support of $f$. 
 
These considerations might suggest that the Garfinkle and Helfer results 
arise by methods of spatial averaging which manage to capture large 
amounts of $(-)$ energy while ignoring larger amounts of $(+)$ energy which 
are really very close by. However, the general result of Helfer is not 
so easily explained. Even if $\Sigma$ is a constant $t$ surface, $\hat{H}$ 
need not be bounded below in general in four-dimensional spacetime. In this 
case, we are dealing with the generalization of the Garfinkle result where 
the walls of the box cease to be sharply defined. This 
seems to suggest that  spatial averaging without time averaging may not be 
sufficient to yield quantities which are bounded below. 
 
\subsection{\label{subsec:spatial:avgeqi}Spacetime averaged quantum 
inequalities in two dimensions} 
 
We now turn to the question of whether one can derive generalizations of 
the quantum inequalities which involve averaging over both space and time. 
In two-dimensional spacetime, this can indeed be done. This was first done by 
one of us \cite{Roman97} using a method analogous to those used in 
Ref.~\cite{F91} to first prove worldline quantum inequalities. 
Flanagan \cite{FLAN} 
later noted that his method may also be used to generate  two-dimensional 
spacetime averaged quantum inequalities. Let $\sigma(u,v)$ be a spacetime 
sampling function, where $u=t-x$ and $v=t+x$ are null coordinates. We will 
assume that this function can be expressed as a product of sampling 
functions in space and time separately in some frame of reference: 
\begin{equation} 
\sigma(u,v) = g_T(t)\, g_S(x) \,. 
\end{equation} 
 The sampled energy density is 
\begin{eqnarray} 
\hat{\rho} &=& \int  T_{tt}  \,g_T(t)\, g_S(x) \, dt\, dx   \nonumber \\ 
&=& \frac{1}{2} \int [T_{uu}(u) + T_{vv}(v)] \, \sigma(u,v)\, du\, dv \,. 
\end{eqnarray} 
Let 
\begin{equation} 
g_1(u) = \frac{1}{2} \int \sigma(u,v)\, dv 
\end{equation} 
and 
\begin{equation} 
g_2(v) = \frac{1}{2} \int \sigma(u,v)\, du \,. 
\end{equation} 
The various sampling functions are normalized so that 
\begin{eqnarray} 
\int \,g_T(t)\, g_S(x) \, dt\, dx &=&  \frac{1}{2} \int \sigma(u,v)\, du\, dv 
      \nonumber \\ 
 = \int g_1(u)\, du \, &=& \int g_2(v)\, dv = 1 \,. 
 \end{eqnarray} 
We can now write the spacetime averaged quantum inequality as 
\begin{eqnarray} 
\hat{\rho} &=&  \int T_{uu}(u)\, g_1(u)\, du + \int T_{vv}(v)\, g_2(v)\, dv 
  \nonumber \\ 
&\geq& - \frac{1}{48\, \pi} \left[ \int_{-\infty}^\infty du\, \frac{(g_1')^2}{g_1} 
+ \int_{-\infty}^\infty dv\, \frac{(g_2')^2}{g_2} \right] \,. 
\end{eqnarray} 
In the last step, we used Flanagan's result, Eq.~(\ref{eq:flanqi}). 
 
Let us explicitly evaluate the bound for some particular sampling functions. 
First consider Lorentzian functions in both space and time, with widths 
$x_0 > 0$ and $t_0 > 0$, respectively: 
\begin{equation} 
g_S(x) = \frac{x_0}{\pi(x^2 +x_0^2)}\, , \qquad 
g_T(t) = \frac{t_0}{\pi(t^2 +t_0^2)}\, . 
\end{equation} 
These choices lead to 
\begin{equation} 
g_1(u) = \frac{x_0+t_0}{\pi[(x_0+t_0)^2 + u^2]} \, , 
\end{equation} 
and $g_2(v) = g_1(v)$. 
The bound on the spacetime averaged energy density now becomes 
\begin{equation} 
\hat{\rho} \geq - \frac{1}{48\, \pi\, (x_0 + t_0)^2} \,. 
                     \label{eq:st_lor} 
\end{equation} 
A second possible choice of sampling function is a Gaussian in both 
space and time: 
\begin{equation} 
g_S(x) = \frac{1}{\sqrt{\pi} x_0}\, {\rm e}^{-x^2/x_0^2}  \, , \qquad 
g_T(t) = \frac{1}{\sqrt{\pi} t_0}\, {\rm e}^{-t^2/t_0^2} \, . 
\end{equation} 
In this case, we find 
\begin{equation} 
g_1(u) = \frac{1}{\sqrt{\pi (x_0^2 + t_0^2)}} \,  {\rm e}^{-u^2/(x_0^2 + t_0^2)}\,, 
\end{equation} 
and the bound becomes 
\begin{equation} 
\hat{\rho} \geq - \frac{1}{12\, \pi\, (x_0^2 + t_0^2)} \,. 
                                       \label{eq:st_gau} 
\end{equation} 
 
These spacetime averaged quantum inequalities reduce to the usual 
QIs along worldlines in the limit that $x_0 = 0$. Note that in 
two dimensions one also has a nontrivial bound from spatial averaging alone 
when $t_0 = 0$. The extent to which the type of results found here in 
two dimensions can be generalized to four dimensions is unclear. It seems 
that there one may need the temporal averaging to get a bound. 
 
\section{\label{sec:fcdistr}Forbidden and constrained distributions} 
 
We can rule out several spatial distributions 
of \p\ and \n\ energy by applying the \AWEC\ and the \QI s 
to their possible evolutions. 
In all of the examples 
given below we assume that the violations of the \WEC\ are strong, 
i.e., if the energy density is negative in one frame then 
it is negative in all frames. (See the Appendix for further details.) 
This assumption is necessary for the following discussion. 
In most of the cases we will be specifically 
considering null fluids, i.e., 
\begin{equation} 
T^{\mu\nu} = \rho \, k^{\mu} k^{\nu} \,. 
\label{eq:nfluid} 
\end{equation} 
For such stress tensors the violations of the \WEC\ are strong, as can be 
easily shown. We will explicitly point out the situations in which we do not 
assume this form for the stress tensor. 
 
\subsection{\label{subsec:fcdistr:f}Forbidden distributions} 
 
\subsubsection{\label{subsubsec:fcdistr:f:sep}Separated 
regions of \p\ and \n\ energy, with the \p\ energy moving rigidly} 
 
Consider an initial state that consists of a compact region, $\cal N$, 
of \n\ energy 
and a distinct compact region, $\cal P$, of \p~energy. 
The compactness of the initial \p\ energy distribution (i.e., 
its finite spatial extent) is crucial to the arguments that 
we present. We assume that $\cal P$ does 
not embrace $\cal N$ in the sense that both regions can be 
contained in non-intersecting rectangular boxes. 
 
We consider situations in which the \p\ energy 
moves ``rigidly'' in that the null flow vector $k^\mu$ in the 
energy-momentum tensor, Eq.~(\ref{eq:nfluid}), is constant in 
Cartesian coordinates.  More general evolutions 
are discussed in Sec.~\ref{subsec:fcdistr:c}. 
The \n\ energy may evolve in any way that it likes. 
 
Let $O$ be any point in $\cal N$ and choose Cartesian coordinates 
with $O$ at the origin. The time axis will be the straight line 
that passes through $O$ in the time direction ($x=y=z=0$). 
Consider the world tube that represents the 
evolution of $\cal P$, extended as far as possible in both 
future and past directions.  If this world tube never crosses 
the time axis, an observer sitting on the axis throughout will never 
encounter the \p\ energy and his worldline will violate the~\AWEC. 
 
\begin{figure} 
\includegraphics[scale=.35]{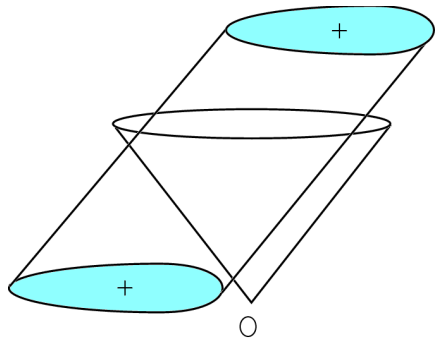} 
\caption{\label{fig:capcone}The world tube of the \p\ energy region cannot 
``cap the future light cone'' of a point, $O$, in the \n\ energy 
region.} 
\end{figure} 
Next, consider the case where the world tube crosses the time axis in 
the future direction. 
(The case when it crosses it in the past direction 
is covered by the time-reverse of the argument presented below.) 
This means that the positive energy flows across 
the future of the region where there was negative energy, as shown in 
Fig.~\ref{fig:capcone}.  Since the motion of the positive energy 
is rigid, its world tube does not expand in either time direction. 
Thus, although it will cross the future light cone of a point in the 
negative energy region, it cannot entirely ``cap'' that light cone. 
Suitably chosen timelike geodesics that go through the negative energy 
can avoid intersecting any positive energy at all.  We prove this 
precisely below. 
 
In the case where the crossing takes place in the future, 
no timelike geodesic through 
$O$ can intersect the world tube of $\cal P$ in the past, since in the 
past direction the 
world tube is running away from the origin at the speed of light. 
We show that there is at least one geodesic that avoids intersecting the 
world tube of $\cal P$ in the future direction as well. 
We choose our Cartesian coordinates so that 
the positive $y$ direction is the direction of motion of the 
\p~energy and we set things up at $t=0$ as follows: 
Let $P=(0,x_0,y_0,z_0)$ be a point in $\cal P$ with the 
property that no point in $\cal P$ has a larger $x$-coordinate. 
The compactness of $\cal P$ guarantees that there will be such a point. 
This point will lie on an ``edge'' of $\cal P$ in the $x$~direction. 
The choice of $x$ is arbitrary. 
We could just as well make the argument we are about to make below by 
choosing a point on the edge in the $z$~direction. 
By the assumptions that 
the \p\ energy does not embrace the \n\ energy and that the \p\ energy 
is moving towards the origin in the positive $y$ direction, we 
have $y_0<0$. 
 
Since the world tube of $\cal P$ moves rigidly in the $y$ direction, 
we note that no point on this world tube can have an $x$~coordinate larger 
than $x_0$.  Our strategy is to show that there are timelike geodesics 
that pass through the origin and escape to a point with $x$~coordinate 
equal to $x_0$ without intersecting the \p~energy (i.e., the world tube 
of $\cal P$). Since the 
$x$~coordinate on such a geodesic must continue to increase, it can 
never intersect the world tube of 
$\cal P$ if it has not already done so by this stage. 
 
\begin{figure} 
\includegraphics[scale=.6]{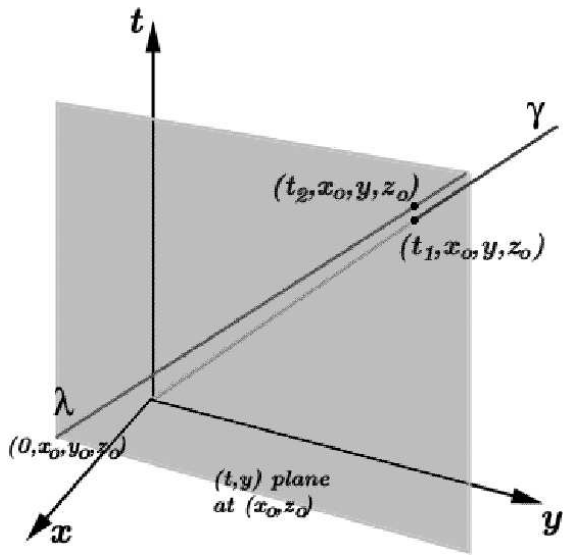} 
\caption{\label{fig:separgu}An illustration of the argument 
of Sec.~\ref{subsubsec:fcdistr:f:sep}. The timelike geodesic $\gamma$ from the 
origin gets to point $(x_0,y,z_0)$ on the $(t,y)$ plane 
at $(x_0,y_0)$, the shaded region, before the null geodesic 
$\lambda$ gets to the same spatial point.} 
\end{figure} 
Let $\lambda$ be the null geodesic in the world tube of $\cal P$ 
that passes through $P$ and points in the direction 
of the flow vector $k^\mu$.  It will obey the equations 
$t=y-y_0, x=x_0, z=z_0$ and so lies in the $t$-$y$ plane located 
at $(x_0,z_0)$, as depicted in Fig.~\ref{fig:separgu}. 
Now consider an arbitrary timelike geodesic, $\gamma$, 
through $O$. It will obey $t^2=\alpha^2(x^2+y^2+z^2)$, where $\alpha>1$. 
If such a $\gamma$ can get to some spatial point $(x_0,y,z_0)$, with $y>0$, 
in the $t$-$y$ plane of interest 
before $\lambda$ gets there, then, as we have seen, $\gamma$ can avoid 
intersecting any positive energy.  We show that it is possible 
for $\gamma$ to do this. 
Suppose that $\gamma$ gets to $(x_0,y,z_0)$ at time $t_1$ and 
$\lambda$ gets there at time $t_2$.  The condition $t_2>t_1$ 
for $\lambda$ to get to this spatial point after 
$\gamma$ can be expressed as 
$(y-y_0)^2 > \alpha^2(x_0^2+y^2+z_0^2)$, or 
\begin{equation} 
\alpha^2 x_0^2 + \alpha^2 z_0^2 <  -y^2 (\alpha^2-1) -2yy_0 +y_0^2. 
\label{eq:time_cond} 
\end{equation} 
Since $y_0<0$ and $y>0$, the only negative term on the right is the first one. 
Choosing $\alpha^2=1+(y_0^2/2y^2)$, we see that (i)~the first and third terms 
on the right combine to give a positive value, and (ii)~for 
large $y$, we have $\alpha^2$ close to~$1$. By choosing $y$ sufficiently large, then, 
we can make the right hand side as 
big as we want while keeping the left hand side close to $x_0^2 + z_0^2$. 
Therefore, irrespective of the values of $x_0$ and $z_0$, we can 
find a value of $y$ (with $\alpha^2$ chosen as above) so as to satisfy 
inequality~(\ref{eq:time_cond}).  Thus, we have shown 
that it is possible to find a timelike geodesic that outruns the 
positive energy to an edge of the spatial region that the positive 
energy can cover.  This geodesic only passes through negative energy~-- 
a forbidden scenario. 
 
The argument covers any finite initial distribution of a \p~energy null fluid, 
no matter how large, and any initial distribution of \n~energy, no matter 
how small, as long as the \p~energy moves rigidly in one direction. 
The \p~energy cannot fully ``cap the future light cone'' of a \n~energy point, as 
illustrated in Fig.~\ref{fig:capcone}. 
Some timelike geodesics are guaranteed to escape without intersecting any 
\p~energy. If we want the AWEC to hold along every timelike geodesic, then 
even if there is a single point at which negative energy exists we need 
an infinitely large distribution of compensating positive energy. 
 
Our argument applies to any shape of \p\ and \n\ energy distribution. 
In particular, it covers \p\ and \n\ energy ``pancakes''.  These 
are distributions that are small in one spatial direction compared 
to the other two. 
 
\subsection{\label{subsec:fcdistr:c}Constrained distributions} 
 
\subsubsection{\label{subsubsec:fcdistr:c:sep}Separated 
regions of \p\ and \n\ energy, with the \p\ energy moving 
arbitrarily} 
 
If the \p\ energy does not move rigidly, the configuration 
discussed in Sec.~\ref{subsubsec:fcdistr:f:sep} cannot be ruled out, 
in general, on the 
grounds that there will always be a timelike geodesic that intersects 
the \n\ energy but not the~\p.  If the \p\ energy expands outward, 
for instance, no timelike geodesic through \n\ can outrun 
the world tube of the \p\ energy.  Even in this case, however, by choosing 
a timelike geodesic that is close enough to a null one we can put 
off the encounter with the positive energy as late as we like. 
If the distribution of \p\ energy is expanding, its density may 
then be dilute enough so as to be insufficient to enforce the \AWEC. 
This will happen in the case when the \p\ energy expands outward 
uniformly, so that its density $\rho$ goes down everywhere as 
$1/t^3$. 
 
\subsubsection{\label{subsubsec:fcdistr:c:pan}Pancakes} 
 
Let us first consider a box-like region of $(-)$ energy which moves 
in the $x$-direction at the speed of light. If the box has a constant 
energy density of magnitude $|\rho|$, how large can the box be? Presumably 
there is some $(+)$ energy nearby, as required by the AWEC and the 
QIs. However, if we use a compactly-supported sampling function 
which cuts off rapidly near the edge of the $(-)$ energy 
region, we can get a bound on the size of the $(-)$ energy region 
alone. 
 
Assume the density, $-|\rho|$, is that measured in the frame of reference 
of an observer $O$ who is at rest on the $x$-axis. We further 
assume that the stress tensor has the null fluid form given by 
Eq.~(\ref{eq:nfluid}), with $\rho = -|\rho|$. Take 
$k^{\mu}=(1,1,0,0)$. The observer's four velocity is 
$u^{\mu}=(1,0,0,0)$. Therefore $\T\mu\nu u^{\mu} u^{\nu}=-|\rho|$. The 
QI applied in $O$'s frame is 
\begin{equation} 
-\int_{-\infty}^{\infty} \, g(t) |\rho| \, dt \geq -\frac{C}{{t_0}^4} \,. 
\label{eq:qio} 
\end{equation} 
Here the Minkowski time $t$ is the proper time along $O$'s worldline 
and $t_0$ is the sampling time, which we set equal to the time $O$ 
spends in the $(-)$ energy region. Let the $x$-dimension of the 
box, as measured by $O$, be $L_x$. Since the box moves past $O$ 
along the $x$-axis at the speed of light, the time for the box to 
pass $O$ is $t_0=L_x$. Using Eq.~(\ref{eq:qio}), and the fact that 
$\rho$ is a constant and the sampling function is compactly-supported 
with unit norm, we obtain the bound 
\begin{equation} 
L_x \leq {\left(\frac{C}{|\rho|}\right)}^{1/4} \,. 
\label{eq:Lbound} 
\end{equation} 
 
Can we get a stronger bound using observers boosted in 
the $x$-direction? First consider an observer $O'$ who is boosted 
along the $x$-axis. His four velocity is 
$u^{\mu}=(\gamma,v \gamma,0,0)$, where $\gamma=1/\sqrt{1-v^2}$, 
and $\T\mu\nu u^{\mu} u^{\nu}=-|\rho| \,(1-v)/(1+v)$. The 
QI applied in $O'$'s frame is 
\begin{equation} 
-\int_{-\infty}^{\infty} \, g(\tau) |\rho| \, 
\left(\frac{1-v}{1+v}\right)\, 
d\tau \geq -\frac{C}{{\tau_0}^4} \,. 
\label{eq:qiob} 
\end{equation} 
Here $\tau$ is $O'$'s proper time coordinate and 
the sampling time $\tau_0$ is the time $O'$ spends in the $(-)$ 
energy. The time $\tau_0$ for the box to pass $O'$ is 
$\tau_0=L_x/\gamma$. Using Eq.~(\ref{eq:qiob}), we get 
the following bound on $L_x$: 
\begin{equation} 
L_x \leq {\left(\frac{C}{|\rho|}\right)}^{1/4} 
\left[\frac{1}{{(1-v)}^{3/4}{(1+v)}^{1/4}}\right]\,. 
\label{eq:Lpbound} 
\end{equation} 
The righthand side has a minimum at $v=-1/2$. The resulting 
bound is 
\begin{equation} 
L_x \leq 0.877 \,{\left(\frac{C}{|\rho|}\right)}^{1/4} \,, 
\label{eq:Lpbound2} 
\end{equation} 
which is slightly stronger than Eq.~(\ref{eq:Lbound}). 
 
Can we constrain the other dimensions of the box by examining 
observers who are boosted in directions transverse to the box's 
direction of motion? It would appear not. Consider an 
observer shot through the box along the $y$-axis. The maximum 
time the observer can spend in the $(-)$ energy is ultimately 
determined by how long the box takes to pass him, which in turn 
depends only on its length along its direction of motion. The 
latter is bounded by Eq.~(\ref{eq:Lpbound2}). Note that the 
length of the $y$-dimension of the box could be as large as 
we like. How far the observer can travel in this direction, 
while still remaining in the $(-)$ energy, depends on how 
long it takes the back wall of the box to hit him. Hence it appears 
that we can make the transverse dimensions of the box as large 
as we like. 
 
The previous discussion leads naturally to a reconsideration 
of ``pancakes'', i.e., ``boxes'' which are much longer in the 
transverse dimensions compared to their thickness in the direction 
of motion. We saw from our earlier discussion that a configuration 
of two finite $(+)$ and $(-)$ energy pancakes was impossible. 
The $(+)$ pancake was required to be of infinite extent in the 
transverse dimensions. There is a further constraint between the 
magnitudes of the relative energy densities,$|\rho_+/\rho_-|$, 
and their separation, $d$, which is given by the quantum 
interest effect \cite{FR99}. Consider a stationary observer who gets 
hit first by the $(-)$ pancake followed by the $(+)$ one. From quantum 
interest we know that the $(+)$ energy density must overcompensate the 
$(-)$ energy density by an amount which grows as the separation $d$ 
increases. 
 
\subsubsection{\label{subsubsec:fcdistr:c:rigid}Rigidly 
moving engulfed \n\ regions} 
Consider a $(-)$ energy region which is enveloped 
by a surrounding $(+)$ energy region. Assume that the shapes 
of the regions are time-independent and that they are null fluids 
which move in one direction. If the energy distributions are 
assumed to be continuous, the boundaries of the worldtubes 
of the $(-)$ and $(+)$ energy must be surfaces of zero energy density. 
Therefore the energy density in each worldtube cannot be constant. 
To satisfy our rigidity requirement, we must have 
$\nabla_{\mu} k^{\mu}=0$; to guarantee energy conservation we must 
have ${T^{\mu\nu}}_{,\nu}=0$. These two criteria will be satisfied, 
with non-constant energy densities, if the densities do not vary along 
the null propagation direction. That is, we assume that 
$k^{\mu} \, \nabla_{\mu} \rho =0$. 
 
\begin{figure} 
\includegraphics[scale=.35]{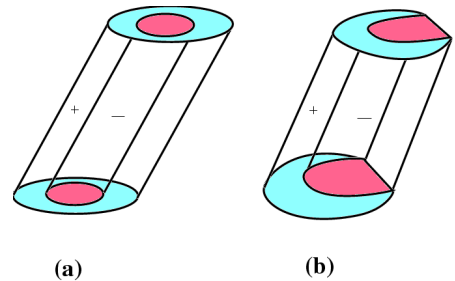} 
\caption{\label{fig:embraced}In $(a)$, the worldtube of a $(-)$ energy region is 
totally surrounded by the worldtube of a $(+)$ energy region. 
In $(b)$, the $(+)$ energy envelops all but the forward-moving 
edge of the $(-)$ energy region.} 
\end{figure}

Any timelike observer who starts in 
the $(-)$ energy region will eventually encounter the $(+)$ energy 
(see Fig.~\ref{fig:embraced}(a)), so this case appears to be allowed. 
However for massive fields, even this 
configuration is impossible, since the two energy regions would 
travel at speeds less than 1. Hence it is always possible to find 
an observer who simply sits in the $(-)$ energy region for an 
arbitrarily long time, which violates the QIs. 
 
Topologically the $(-)$ energy region here 
is equivalent to the $(-)$ energy ``box'' discussed earlier in 
this section. Therefore, for the null fluid case we can use the 
same argument to place constraints on the magnitude of the 
$(-)$ energy density in the interior region and the 
thickness in its direction of motion. However, here the $(+)$ 
energy actually need envelop only all but a line of 
tangency which is transverse to the direction of motion, as depicted in 
Fig.~\ref{fig:embraced}(b). In this case as well, 
there are no timelike observers who intersect {\it only} 
the $(-)$ energy. As an aside, we point out that the limiting 
case is when the line of tangency is shrunk to a point 
which lies along the direction of motion. 
 
\subsubsection{\label{subsubsec:fcdistr:f:eeshells}Expanding engulfed \n\ 
energy shells} 
 
\begin{figure} 
\includegraphics[scale=.35]{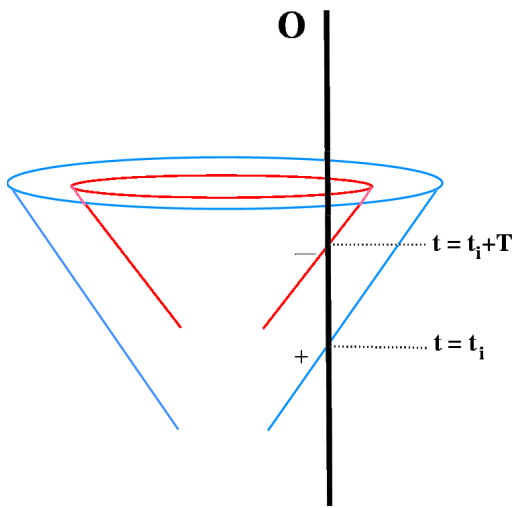} 
\caption{\label{fig:nestcones} Two spatially 
concentric, radially expanding $\delta$-function null shells 
of $(+)$ and $(-)$ energy, created in the past. 
The worldline of observer $O$ intersects the $(+)$ energy 
shell at time $t=t_i$, and the $(-)$ energy shell a time $T$ later.  
The vertices of the two light cones, although not shown, are assumed  
to lie along the same line, which is parallel to $O$.} 
\end{figure} 
 
Consider two spatially concentric, radially expanding  
$\delta$-function null shells 
of $(+)$ and $(-)$ energy, which were created at two 
different times in the past. 
A stationary observer $O$ is hit first by the $(+)$ shell at $t=t_i$, 
and later by the $(-)$ shell at $t=t_i+T$, where $T$ is the separation 
between the shells in time. This scenario is depicted in 
Fig.~\ref{fig:nestcones}. Let the energy density of the $(+)$ 
shell, as measured by $O$ at time $t_i$, be 
\begin{equation} 
\rho_+ = \frac{a}{t_i^2} \, \delta(t-t_i) \,, 
\label{eq:rho+} 
\end{equation} 
with $a=\,{\rm constant}\,>0$, and the energy density at 
$t=t_i+T$ be 
\begin{equation} 
\rho_- = -\frac{b}{t_i^2} \, \delta(t-(t_i+T)) \,, 
\label{eq:rho-} 
\end{equation} 
with $b=\,{\rm constant}\,>0$. The constants $a$ and $b$ are 
measures of the magnitudes of the energy densities, 
neglecting the effects of expansion. This scenario can be 
constrained using the QI's with a compactly supported sampling function, 
following the argument given in Sec.~III of \cite{FR99}. 
 
Choose a compactly supported sampling function with a single maximum 
centered on $t=t_i+T$ (i.e., on the $(-)$ energy shell), with a 
width $t_0$. Substituting Eqs.~(\ref{eq:rho+}) and ~(\ref{eq:rho-}) 
into the QI, we get 
\begin{eqnarray} 
\hat \rho &=& \int_{-\infty}^{\infty} \, g(t) \, \rho (t) \, dt 
 \nonumber \\ 
&=& \frac{a}{t_i^2} \, g(t_i) - \frac{b}{t_i^2} \, g(t_i+T) 
 \, \geq -\frac{C}{{t_0}^4} \,. 
\label{eq:genrho} 
\end{eqnarray} 
If we now choose the width of the sampling function to be 
$t_0=2T$, then $g(t_i)=0$. For a sampling function with one 
maximum at $g(t_i+T)$, $g(t_i+T) \propto 1/{t_0}$, so let 
$g(t_i+T) = C_0/{t_0}=C_0/2T $, where $C_0$ is a constant 
whose value depends only on the form of the chosen sampling 
function (but not on the spacetime dimension, unlike $C$). 
Therefore we obtain 
\begin{equation} 
T \leq {\left( \frac{C}{8 b\,C_0} \right)}^{1/3} \, {t_i}^{2/3} \,. 
\label{eq:T3} 
\end{equation} 
We see that for fixed $t_i$, $T$ decreases with increasing $b$, as expected. 
When $b$ is fixed and for $0<t_i \ll T$, we see that $T$ must decrease as  
$t_i$ decreases. (To avoid singularities in the energy 
densities we do not want to allow $t_i \rightarrow 0$, which is why  
only the later stages of the evolution are illustrated in  
Fig.~\ref{fig:nestcones}.) In the limit when $t_i \gg T$, for fixed $b$,  
the bound Eq.~(\ref{eq:T3}) becomes very weak. 
 
\subsubsection{\label{sec:fcdistr:c:sshells}Separated Expanding Shells of \p\ 
  and \n\ Energy} 

\begin{figure} 
\includegraphics[scale=.35]{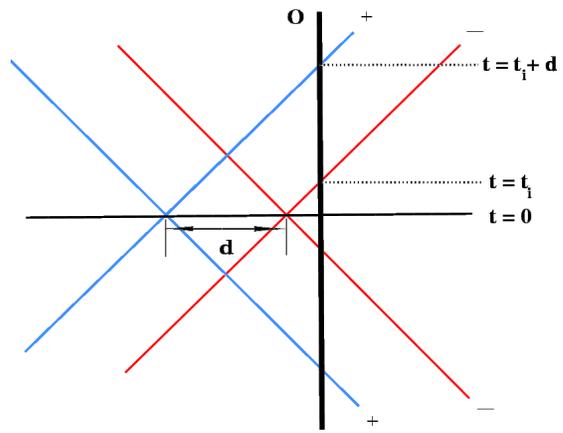} 
\caption{\label{fig:sepcones}Two contracting shells of $(+)$ and $(-)$ energy 
reach their maximum densities at $t=0$, and subsequently 
re-expand. The regions of maximum density are separated by 
distance $d$. A stationary observer $O$ intersects the $(-)$ 
energy shell at $t=t_i$.} 
\end{figure} 
Consider two null fluid $\delta$-function shells of separated $(+)$ and 
$(-)$ energy which contract and re-expand. The shells reach 
maximum density at time $t=0$. The spatial locations 
where the densities become maximum are separated by a distance, 
$d$. (We ignore any interactions when the shells cross each other.) 
The evolution of the shells is depicted 
in Fig.~\ref{fig:sepcones}. A static observer, $O$, gets 
hit by each shell twice. The worldline of $O$ 
crosses the $(-)$ energy shell for a second time at time $t=t_i>0$, 
and crosses the $(+)$ energy shell for the second time at time 
$t=t_i+d$. Note that the diagram is time-symmetric around $t=0$. 
 
The following argument uses only the AWEC to constrain this scenario. 
As before, let the magnitudes of the energy densities 
(neglecting the effects of contraction and expansion) 
be ``$a$'' for the $(+)$ 
energy shell and ``$b$'' for the $(-)$ energy shell, with 
$a,b$ chosen to be positive constants. If we apply the AWEC 
to $O$'s worldline, and use the time-symmetry of the diagram, 
we obtain 
\begin{equation} 
\int_{-\infty}^{\infty} \, T_{\mu\nu}u^{\mu}u^{\nu} \, dt 
= \frac{2 a}{{(t_i+d)}^2} - \frac{2 b}{t_i^2} \, \geq 0 \,. 
\label{eq:intecones} 
\end{equation} 
The factors of $2$ reflect the fact that $O$ gets hit by 
each shell twice. If we let $f=a/b$, we can rewrite this as 
\begin{equation} 
f t_i^2 \geq {(t_i+d)}^2 \,. 
\label{eq:sepbound} 
\end{equation} 
The quantities $f$, $t_i$, and $d$ are all positive, so we obtain  
\begin{equation} 
d \leq ( {\sqrt f} - 1) \, t_i \,. 
\label{eq:dbound} 
\end{equation} 
This implies that $f>1$, and that $f$ must increase as $t_i/d$ decreases. 
In the limit $t_i/d \gg 1$, we simply get $f \agt 1$, which is a 
fairly weak bound. 
 
The bound Eq.~(\ref{eq:dbound}) becomes more and more stringent as $t_i$ 
decreases. However, it is more realistic to suppose that the shells have a finite 
thickness $\Delta$. This can be viewed as either the thickness in space at 
a fixed time, or else the duration in time along $O$'s worldline. 
Then the above analysis holds so long as $t_i > \Delta$, 
and the best bound, obtained when $t_i \approx \Delta$, implies that 
\begin{equation} 
f \agt \frac{(d + \Delta)^2}{\Delta^2} \,. 
\label{eq:fbound} 
\end{equation} 
When $d \gg \Delta$, this requires $f \gg 1$, which is a version of the 
quantum interest phenomenon. 
 
\section{\label{sec:explicit}Explicit construction of allowed distributions} 
 
\subsection{Plane Wave Modes} 
\label{sec:pwmodes} 
In the previous section, we discussed $(-)$ energy 
distributions which were either ruled out or constrained 
by the AWEC and the QIs. We now give some examples of 
distributions which can be explicitly constructed from 
allowed states in quantum field theory, and analyze some 
of their properties. The class of examples which we will 
focus on are squeezed vacuum states, which are discussed 
extensively in quantum optics and which can now be 
constructed in the laboratory \cite{SZ}. Our discussion 
is restricted here to quantized massless and massive minimally 
coupled scalar fields in flat spacetime, but it could be easily 
generalized to include the electromagnetic field as well, 
which is also known to obey the QIs and the AWEC \cite{FR97,P01}. 
The stress tensor for the minimally coupled scalar field is 
\begin{equation} 
T_{\mu \nu} = \phi_{,\mu}\phi_{,\nu} 
-{1\over 2}\, \eta_{\mu \nu} \, \left(\phi_{,\alpha}\phi^{,\alpha}\, 
+m^2 \, \phi^2 \right) \,. 
\label{eq:scalarst} 
\end{equation} 
The field operator may be expanded in terms of creation and annihilation 
operators as 
\begin{equation} 
\phi = {\sum_k} ({a_k}{f_k} + {a_k}^\dagger{f_k}^\ast) \,. 
\label{eq:phiop} 
\end{equation} 
For simplicity, we will consider only a single mode state 
with $t$ and $x$ dependence only, i.e., $f_k=f=f(t,x)$. 
The renormalized expectation values of the energy density, 
pressure, and flux are then given by 
\begin{eqnarray} 
T_{00} &=& Re \, \left[\langle {a^\dagger a} \rangle \, 
(f^*_{,t} \,f_{,t} + f^*_{,x} \,f_{,x}) + \langle {a^2}\rangle \, 
({f_{,t}}^2 + {f_{,x}}^2) \right. \nonumber \\ 
&&+ \left. m^2 \,(\langle {a^\dagger a} \rangle 
\,f^* f + \langle {a^2}\rangle \,f^2) \right] \,, 
\label{eq:expenergy} \\ 
T_{11} &=& Re \, \left[\langle {a^\dagger a} \rangle \, 
(f^*_{,x} \,f_{,x} + f^*_{,t} \,f_{,t}) + \langle {a^2}\rangle \, 
({f_{,x}}^2 + {f_{,t}}^2) \right. \nonumber \\ 
 &&- \left. m^2 \,(\langle {a^\dagger a} \rangle 
\,f^* f + \langle {a^2}\rangle \,f^2) \right] \,, 
\label{eq:exppress} \\ 
T_{01} &=& Re \, \left[\langle {a^\dagger a} \rangle \, 
(f^*_{,t} \,f_{,x} + f^*_{,x} \,f_{,t}) + 2\langle {a^2}\rangle \, 
(f_{,t} \,f_{,x})  \right] \!\!, 
\label{eq:expflux} 
\end{eqnarray} 
respectively. 
 
Here the mode function will be taken to be a 
plane wave mode of the form 
\begin{equation} 
f= {i \over \sqrt{2\omega L}} e^{i(kx - \omega t)}, 
\label{eq:fplane} 
\end{equation} 
with $\omega=\sqrt{k^2+m^2}$, and where 
${\bf k}=k {\bf \hat x}$,  and 
a periodicity of length $L$ has been imposed in the 
spatial direction, so that $k$ takes 
on discrete values. We choose the quantum state 
$|\psi \rangle$ to be a squeezed vacuum state: 
\begin{equation} 
|\xi \rangle = S(\xi)|0 \rangle \,, 
\label{eq:sqstate} 
\end{equation} 
 where $S(\xi)$ is the ``squeeze operator,'' and 
$\xi=r \, e^{i \theta}$ is an arbitrary complex number. 
In this state, 
\begin{equation} 
\langle {a^\dagger a} \rangle = {{\rm sinh}^2}r \,, 
\label{eq:sqada} 
\end{equation} 
and 
\begin{equation} 
\langle {a^2} \rangle = -{\rm sinh}r \, {\rm cosh}r \,, 
\label{eq:sqaa} 
\end{equation} 
where $r > 0$ is the squeeze parameter, and where we have 
chosen the phase $\theta=0$ \cite{SZ}. 
If we substitute Eqs.~(\ref{eq:fplane})-~(\ref{eq:sqaa}) 
into Eqs.~(\ref{eq:expenergy})-(\ref{eq:expflux}), we obtain 
\begin{eqnarray} 
\!\!\!\!\!\!T_{00} &\!\!\!=\!\!\!& {\frac{\omega}{L}} \, {\rm sinh}r 
\left[{\rm sinh}r-{\frac{k^2}{\omega^2}} \, 
{\rm cosh}r \, {\rm cos} \,2(kx-\omega t) \right] \!\!\! \,, 
\label{eq:sqenergy} \\ 
\!\!\!\!\!\!T_{11} &\!\!\!=\!\!\!& {\frac{\omega}{L}} \, {\rm sinh}r 
\left[{\frac{k^2}{\omega^2}}\,{\rm sinh}r 
-{\rm cosh}r \, {\rm cos} \,2(kx-\omega t) \right] \!\!\! \,, 
\label{eq:sqpress} \\ 
\!\!\!T_{01} &=&-{\frac{k}{L}} \, {\rm sinh}r 
\left[{\rm sinh}r-{\rm cosh}r \, 
{\rm cos} \,2(kx-\omega t) \right]  \,. 
\label{eq:sqflux} 
\end{eqnarray} 
\begin{figure} 
\includegraphics[scale=.35]{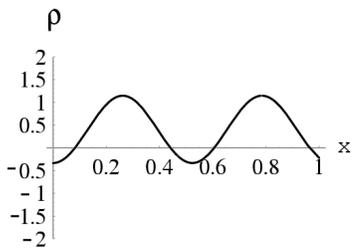} 
\caption{\label{fig:sqenergy}The energy density $\rho$ in a single 
plane wave mode squeezed vacuum 
state, at constant time $t=0$, as a function of position, $x$. 
Here $r= 0.2$, $\omega = 10$, and $m = 8$.} 
\end{figure} 
The energy density as a function of position at fixed 
time is plotted in Fig.~\ref{fig:sqenergy}. One obtains a 
similar graph of energy density as a function of time 
at fixed position. The energy density oscillates between 
$(+)$ and $(-)$ values, with the $(+)$ energy always 
overcompensating the $(-)$ energy. 
 
\begin{figure} 
\includegraphics[scale=.35]{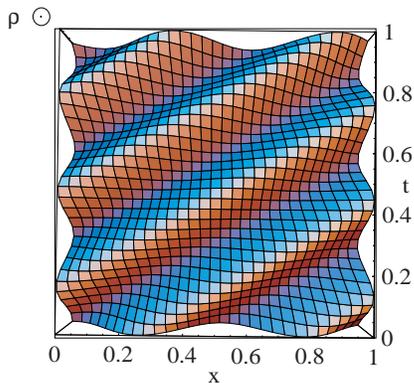} 
\caption{\label{fig:3denergy}A ``top down'' view of the energy density, $\rho$ in a 
single plane wave mode squeezed vacuum state of a massive scalar field, as 
a function of $x$ and $t$. The energy density increases in the direction 
perpendicular to the page. The $(-)$ energy is concentrated along 
spacelike regions. Again $r= 0.2$, $\omega = 10$, and $m = 8$.} 
\end{figure} 
 
\begin{figure} 
\includegraphics[scale=.77]{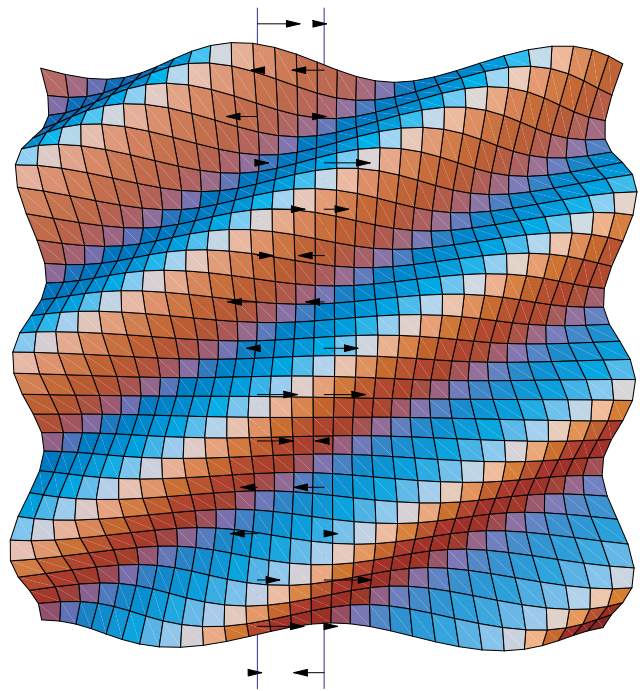} 
\caption{\label{fig:fluxenergy}A ``top down'' view of the energy density, $\rho$ in a 
single plane wave mode squeezed vacuum state of a massive scalar field, 
as a function of position (horizontal axis) and time (vertical axis). 
The energy density increases in the direction out of the page. 
Null lines are at $45$ degrees. 
The $(-)$ energy is concentrated along 
spacelike regions. The arrows indicate the instantaneous direction 
of flow of the energy.  Again $r= 0.2$, $\omega = 10$, and $m = 8$.} 
\end{figure}

For the massive field case it might seem that an 
observer could ride along with the $(-)$ energy in 
violation of the QIs. However, since the 
QIs hold for {\it all} quantum states in flat spacetime, 
we know this cannot be possible. How is this apparent 
paradox resolved? The energy density as a function 
of $t$ and $x$ is plotted in Fig.~\ref{fig:3denergy}. 
The $(-)$ energy density is concentrated along spacelike 
regions. So an observer cannot ride along with it. 
It might appear from this example that the $(-)$ energy is 
``propagating'' along spacelike trajectories. However, 
a relativistic quantum field theory incorporates causality 
in its construction. So what is going on? One must remember 
that these are rather special states which have correlations 
built into them. These built-in correlations cause the energy 
density to vary in a manner that looks like acausal 
propagation. At each point, the energy density is moving in such a way as 
to create the effect of peaks and troughs of energy that are constant 
along spacelike lines. This is illustrated in Fig.~\ref{fig:fluxenergy}. 
A useful analogy is the following. Imagine an 
system of light bulbs with triggering mechanisms and clocks 
which are arranged in a line. An observer can pre-program 
each bulb to be triggered at a certain time. This can be 
done in such a way that another observer who 
later sees the succession of flashes, and interprets them as 
causally generating one another, will think that the flashes 
are propagating faster than light. The correlations of the 
flash times of the bulbs relative to one another have 
been causally pre-programmed into the state of the system from the 
beginning. Another analogy is an Einstein-Podolsky-Rosen state 
in which two photons are generated in an entangled state such 
that a measurement of the spin of one photon allows one to 
determine the spin of the other photon even at spacelike 
separations. This process cannot be used for superluminal 
signaling because there is no way to know ahead of time 
what the spin of the first photon will be before it is measured, 
which is what one would need to send Morse-code type messages. 
The two photons are in some sense two parts of one 
single ``object''. 
 
Another apparent paradox looms at this point. If the $(-)$ 
energy is concentrated along spacelike lines, as shown in the 
figures, then it would seem that a suitably boosted observer 
could make one of these lines a constant time surface on which 
the energy density is everywhere negative. However, these surfaces 
are perpendicular to the observer's timelike Killing vector 
(unlike the spacelike surfaces discussed in Fig.~\ref{fig:helf_slice}), 
and so we know that the energy density 
integrated over all space must be positive. 
 
In the boosted frame, 
\begin{equation} 
T'_{00}= \gamma^2 \, T_{00} + 2 v \, \gamma^2 \, T_{01} 
+ v^2 \, \gamma^2 \, T_{11} \,, 
\label{eq:eboost} 
\end{equation} 
and 
\begin{equation} 
T'_{01}= -v \, \gamma^2 \,(T_{00} + T_{11}) 
-\gamma^2 \, (1 + v^2) \, T_{01} \,. 
\label{eq:fboost} 
\end{equation} 
In the boosted frame described above, $T'_{01}=0$. 
A short calculation shows that this is the case when 
$v=k/\omega$. It is easily shown that this is the 
value of $v$ which gives ${\bf k}'=0$. In this frame 
\begin{equation} 
T'_{00}=\frac{m^2 \, {{\rm sinh}^2}r}{\omega L} 
= {\rm constant} \, >0 \,, 
\label{eq:toop} 
\end{equation} 
so the observer simply sees a constant 
$(+)$ energy density. This is consistent with the fact 
that the WEC violations here are weak. We can show 
this more generally using the results of the Appendix, 
as follows. 
 
For the massive scalar field in a plane wave squeezed vacuum state, 
when $T_{00}<0$, what are the conditions that in a boosted 
frame $T'_{00}<0$ as well? Let $\kappa=kx-\omega t$. 
Since $r>0$ and $\omega > k$, we 
need $(\omega^2/k^2)\,{\rm tanh}r < {\rm cos}\,2\kappa$ 
for $T_{00}<0$, which in turn implies that 
$T_{11}<0$ and $T_{01}>0$, from 
Eqs.~(\ref{eq:sqenergy})-(\ref{eq:sqflux}). 
Therefore we may write 
\begin{eqnarray} 
|T_{11}|&=&-T_{11} = {\frac{\omega}{L}} \, {\rm sinh^2}r 
\left[{\rm coth}r \,\, {\rm cos}\, 2 \kappa 
{-\frac{k^2}{\omega^2}}\right] \,, 
\label{eq:asqpress} \\ 
|T_{01}|&=&T_{01} = {\frac{k}{L}} \, {\rm sinh^2}r 
\left[{\rm coth}r \,\, {\rm cos}\, 2 \kappa -1 \right] \,. 
\label{eq:asqflux} 
\end{eqnarray} 
Note that \mbox{${\rm coth}r \,\,{\rm cos}\, 2 \kappa -({k^2}/{\omega^2}) 
>{\rm coth}r \,\, {\rm cos}\, 2 \kappa -1 
> 0$}, and $(\omega/L) \, {\rm sinh^2}r 
> (k/L) \, {\rm sinh^2}r$, so we have that 
\mbox{$|T_{11}|> |T_{01}|$} and $T_{11}<0$. Thus we have 
an example of Case $1$ of the Appendix, where 
the necessary and sufficient condition for a strong 
WEC violation is Eq.~(\ref{eq:pita}), 
\begin{equation} 
T_{00}<\frac{{T_{01}}^2}{T_{11}} \,. 
\end{equation} 
Since $T_{11}<0$, this implies $T_{00} T_{11} >{T_{01}}^2$. 
Combining Eqs.~(\ref{eq:sqenergy})-(\ref{eq:sqflux}), we find 
\begin{equation} 
T_{00} T_{11} -{T_{01}}^2= 
-(m^2/{\omega^2 L^2})\,{\rm cosh}r\,\, 
 {\rm sinh^3}r \,\, {\rm cos}\, 2 \kappa < 0 \,, 
\label{eq:violation} 
\end{equation} 
since if $T_{00}<0$, then ${\rm cos}\, 2 \kappa > 0$. 
Hence the condition, Eq.~(\ref{eq:pita}), is violated 
and the WEC violation by the massive scalar field in the 
single plane wave mode squeezed vacuum state is weak. 
 
For the massless scalar field, $m=0$ and hence 
$\omega=|k|$,  $T_{00}=T_{11}=-T_{01}$, so this 
is an example of Case 2.2 of the Appendix, 
for which the necessary and sufficient condition 
for strong WEC violation is Eq.~(\ref{eq:nec_cond}), 
\begin{equation} 
|T_{00}| \geq T_{11} + 2 |T_{01}| \,, 
\end{equation} 
which is marginally satisfied in this case. Hence 
for the massless scalar field the WEC violation 
is strong. 
 
\subsection{Wavepackets} 
\label{sec:wpack} 
We now analyze the distribution of $(+)$ and $(-)$ energy 
in a wavepacket of the massive scalar field in 
two-dimensional spacetime. In the mode expansion 
of the field operator, given in Eq.~(\ref{eq:phiop}), we 
will take the $f_k$'s to be a complete orthonormal set 
of wavepacket modes. Let only one single wavepacket 
mode, $f$, be excited, and take the form of this mode to be 
\cite{Gas-1st} 
\begin{eqnarray} 
f(x,t)&=&\frac{\alpha^{1/4}}{2^{3/4} \, 
\pi^{1/4} \, \sqrt{\omega_0}} \, 
{\left(1+\frac{\sigma}{4 \, \alpha\, \omega_0}\right)}^{-1/2} 
\nonumber \\ 
&\!& \times 
e^{i\,(k_0 \,x - \omega_0 \,t)} \, 
e^{-{{(x-v_g \,t)}^2}/4 \,(\alpha+ i\, \sigma\,t)} \,, 
\label{eq:wp} 
\end{eqnarray} 
where $\omega_0 = \sqrt{k_0^2 +m^2}$, 
\begin{equation} 
v_g={\left(\frac{d\omega}{dk}\right)}_{k_0} = \frac{k_0}{\omega_0} \,, 
\label{eq:vg_def} 
\end{equation} 
is the group velocity of the packet, and where 
\begin{equation} 
\sigma=\frac{1}{2}\,{\left(\frac{d^2\omega}{dk^2}\right)}_{k_0} 
      = \frac{m^2}{2 \omega_0^3} \,. 
\label{eq:sigma_def} 
\end{equation} 
\begin{figure} 
\includegraphics[scale=.95]{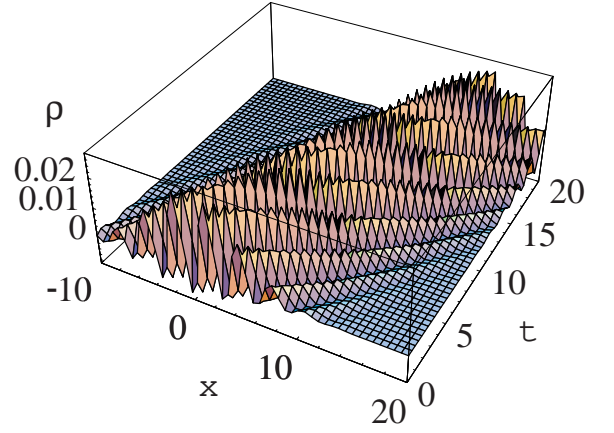} 
\caption{\label{fig:wp}The energy density $\rho$ as a function of $x$ and $t$ 
 for a massive scalar field in a wavepacket mode squeezed vacuum state. 
The peak of the packet follows a timelike trajectory. The negative 
energy is concentrated along the spacelike troughs. Here $r=0.2$, $k_0=0.6$, 
$m=8$ and $\alpha = 20/k_0^2$.} 
\end{figure} 
The packet is sharply peaked around $k_0$ in momentum space 
with spread $1/\sqrt{\alpha}$, where we assume that $\alpha \gg 1$. 
With these assumptions the wavepacket has unit Klein-Gordon norm. 
As before take the quantum state to be a squeezed vacuum state, and 
substitute Eq.~(\ref{eq:wp}) into Eq.~(\ref{eq:expenergy}). A tedious 
calculation then yields a rather long expression for $T_{00}$ 
which we do not reproduce here. A plot of $\rho=T_{00}$ as a function 
of $t$ and $x$ is shown in Fig.~\ref{fig:wp}. Note that the peak 
of the wavepacket moves along a timelike trajectory with the group 
velocity, $d\omega/dk$, whereas the individual components move with 
the phase velocity, $\omega/k$. As in the plane wave case for the 
massive scalar field, the negative energy is concentrated along 
spacelike regions. 
 
\section{\label{sec:awec}The \AWEC\ along geodesic segments} 
 
In this section, we will depart somewhat from the principal topic of this paper 
and discuss some of the limitations of the AWEC. We illustrate why the AWEC 
integral must be taken along a complete geodesic path. We will also have an 
opportunity to explore examples of both strong and weak violations of the 
WEC. 
 
\subsection{A Counterexample to the AWEC for Piecewise Geodesics} 
 
\begin{figure} 
\includegraphics[scale=.35]{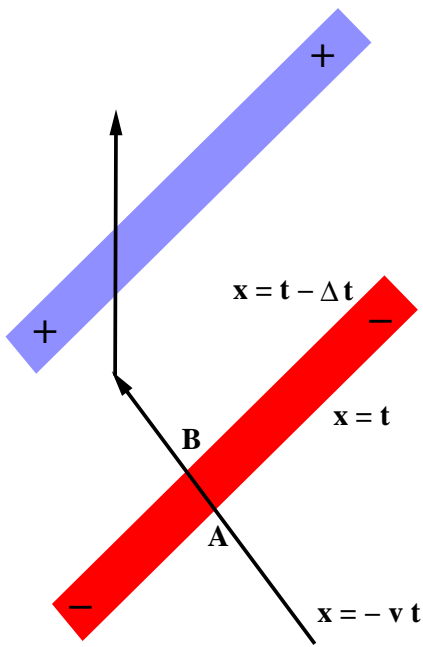} 
\caption{\label{fig:piecewise} An observer on a piecewise geodesic 
path moves through separated 
regions of $(+)$ and $(-)$ energy. The observer moves through the 
$(-)$ energy on the path $ x = - v t$. The lower boundary of the $(-)$ energy 
region is the line $x = t$ and its upper boundary is the line $x= t - \Delta t$.} 
\end{figure} 
 
In this subsection we wish to show that the averaged weak energy condition 
does not hold, even in Minkowski spacetime, if one integrates the 
energy density along a piecewise geodesic path, as opposed to a true geodesic. 
Consider an energy distribution with the null fluid form for the stress 
tensor, Eq.~(\ref{eq:nfluid}). 
Suppose that there are separated $(+)$ and $(-)$ energy regions, both 
moving to the right, as illustrated in Fig.~\ref{fig:piecewise}. 
For the purposes of our 
example, we may take the energy density to be constant within each region, 
so that $\rho = \rho_+$ in the $(+)$ energy region and $\rho = \rho_-$ in 
the $(-)$ energy region. Further require that both pulses last for the same 
time interval $\Delta t$ as measured in the laboratory frame. This means 
that we must have 
\begin{equation} 
\rho_+ > |\rho_-| 
\end{equation} 
in order that there be net $(+)$ energy. 
 
Now consider an observer moving to the left with speed $v$, and hence with 
four-velocity $u^\mu = \gamma (1,-v,0,0)$, where $\gamma = 1/\sqrt{1-v^2}$. 
The energy density in the frame of this observer is 
\begin{equation} 
T^{\mu\nu} u_\mu u_\nu = \rho\, \gamma^2\, (1+v)^2 \,. 
\end{equation} 
Further suppose that this observer moves along the piecewise geodesic 
worldline depicted in Fig.~\ref{fig:piecewise}. The observer first moves at speed $v$ 
through $(-)$ energy, and then is at rest when the $(+)$ energy passes 
by. The path of the observer in the $(-)$ energy can be taken to be 
given by $x = -v t$, and the boundaries of the $(-)$ energy to be the 
lines $x = t$ and $x = t - \Delta t$. The observer enters the \n\  energy 
at point $A$, where $x=t=0$. Let $T$ be the coordinate time 
required to traverse the $(-)$ energy region. At point $B$, we have 
$x = -v T = T - \Delta t$. Hence the proper time which the observer spends 
in the $(-)$ energy is 
\begin{equation} 
\tau = \frac{T}{\gamma} = \frac{\Delta t}{\gamma (1 + v)} \,. 
\end{equation} 
The integrated energy density along this observer's worldline is 
\begin{eqnarray} 
\int T^{\mu\nu} u_\mu u_\nu d\tau 
&=& \rho_+ \, \Delta t + \rho_- \, \tau \,\gamma^2\, (1+v)^2 
                                                     \nonumber \\ 
 &=& \Delta t \left(\rho_+ - |\rho_-| \, \sqrt{\frac{1+v}{1-v}} \right) \,. 
\end{eqnarray} 
So long as $\rho_- \not= 0$, we can find a $v$ which makes this 
expression negative. The piecewise nature of the worldline allows the 
 $(-)$ energy to be enhanced in magnitude by the Doppler shift factor 
$\sqrt{(1+v)/(1-v)}$, while the $(+)$ energy is unchanged.

\subsection{Violations of the Difference AWEC} 
 
The AWEC in its simple form need not hold inside of a cavity, 
if there is negative 
Casimir energy density. In this case, an observer can sit in constant negative 
energy density for an infinite amount of proper time. However, the difference 
between the energy density in an arbitrary quantum state and in the Casimir 
vacuum does satisfy the AWEC. More generally, this difference satisfies quantum 
inequalities, as was discussed in Ref.~\cite{FR95,Mitch98}. These ``difference 
inequalities'' reduce to the ``difference AWEC'' in the limit of long sampling 
times. The latter is the statement that the integral of the difference in energy 
densities is non-negative when integrated over the worldline of an observer 
at rest within the cavity. However, just as it is possible to temporarily 
suppress the local energy density below zero in empty Minkowski spacetime, 
it is possible to find quantum states in which the local energy density 
is more negative than in the Casimir vacuum state. The question which we wish 
to address  in this subsection is the following: Is it possible for a 
moving observer to pass through a cavity in such a way as to see a net negative 
integrated energy from a quantum field confined within the cavity? 
Here we are concerned only with the stress tensor of the quantum field, 
and are ignoring any contributions from the walls of the cavity itself. 
 
Consider a massless scalar field confined between reflecting boundaries located 
at $x = 0$ and at $ x = L$, and a geodesic observer moving at speed $v$ 
in the positive $x$-direction.  The four velocity 
of the observer is $u^\mu = \gamma (1,v,0,0)$ and the energy density in 
this observer's rest frame is 
\begin{equation} 
\rho = T^{\mu\nu}\, u_\mu\, u_\nu \,. 
\end{equation} 
Here $T^{\mu\nu}$ is understood to be the difference in the expectation 
values of the stress tensor operator in a given quantum state and in the 
Casimir vacuum. Let the quantum state be one in which a single mode, 
with mode function $f$, is excited. Then the components of $T^{\mu\nu}$ 
are given by the same expressions that hold for the normal-ordered stress 
tensor in Minkowski spacetime, namely Eqs.~(\ref{eq:expenergy}), 
~(\ref{eq:exppress}), and ~(\ref{eq:expflux}). We may use these expressions 
to write $\rho$ as 
\begin{eqnarray} 
\rho &=& 2 {\rm Re} \{(\gamma^2 - \frac{1}{2})\, 
[|f_{,t}|^2 \langle a^\dagger a \rangle + (f_{,t})^2 \langle a^2 \rangle] 
                                                \nonumber \\ 
&+& (v^2 \gamma^2 + \frac{1}{2})\, 
[|f_{,x}|^2 \langle a^\dagger a \rangle + (f_{,x})^2 \langle a^2 \rangle] 
                                                \nonumber \\ 
&+& v \gamma^2 [(f_{,t}^* f_{,x}+ f_{,x}^* f_{,t})\langle a^\dagger a \rangle 
 +2 f_{,t} f_{,x} \langle a^2 \rangle ] \} \, .  \label{eq:cavity_rho} 
\end{eqnarray} 
We take the mode function to be that of a standing wave which vanishes on the 
walls of the cavity and has no dependence upon the transverse directions: 
\begin{equation} 
f = f(t,x) = A\, \sin \omega x \, {\rm e}^{-i \omega t} \,. 
                                              \label{eq:cavity_mode} 
\end{equation} 
Note that the standing wave modes must satisfy 
\begin{equation} 
\omega = \frac{\pi n}{L} \, , \quad n = 1, 2, 3, \cdots \,. 
\end{equation} 
 
We wish to examine the integrated energy density along this observer's worldline. 
Here it is assumed that there are no particles outside of the cavity, so 
the difference in energy densities is nonzero only inside of the cavity. 
The integrated energy density difference then becomes 
\begin{equation} 
{\cal E} = \int \rho \, d\tau = \frac{1}{\gamma}\, 
           \int_{t_0}^{t_0 + \Delta t} \rho\, d t \,, 
\end{equation} 
where $\Delta t = L/v$ is the coordinate time required to traverse the cavity, 
and $t = t_0$ is the time at which the observer enters. 
Let the quantum state be the single mode squeezed vacuum state discussed 
in Sec.~\ref{sec:explicit}. We can now use Eqs.~(\ref{eq:cavity_rho}), 
~(\ref{eq:cavity_mode}), ~(\ref{eq:sqada}), and ~(\ref{eq:sqaa}) to 
write 
\begin{eqnarray} 
{\cal E} &=&  \frac{A^2\, \omega \, \sinh r}{4\, \sqrt{1-v^2}} \;  \Biggl\{ 
\cosh r \, \biggl[ 2 \sin(2 \omega t_0)  \nonumber \\ 
 &-& (1+v) \sin \{2 [(1+v) \Delta t +t_0]\omega\}  \nonumber \\ 
  &-& (1-v) \sin \{2 [(1-v) \Delta t +t_0]\omega\} \biggr] \nonumber \\ 
  &+& 4 (1+v^2)\, \omega \,\Delta t \, \sinh r \Biggr\} \,. 
\end{eqnarray} 
 
\begin{figure} 
\includegraphics[scale=.5]{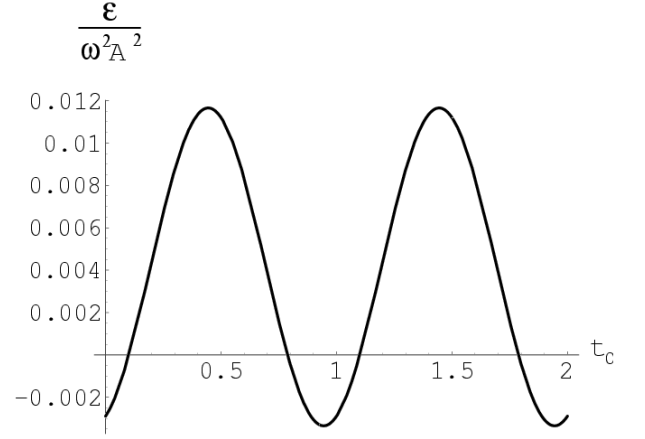} 
\caption{\label{fig:evt} ${\cal E}/\omega^2 A^2$ is plotted as a function of the 
entrance time $t_0$ for the case that $r = 0.03$ and $v = 0.9$. } 
\end{figure} 
\begin{figure} 
\includegraphics[scale=.5]{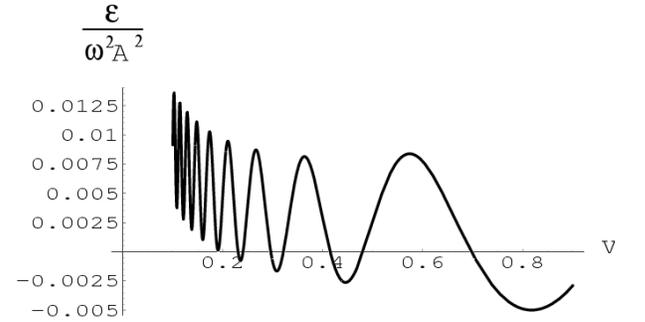} 
\caption{\label{fig:evs}  ${\cal E}/\omega^2 A^2$ is plotted as a function 
of the  observer's speed $v$ for the case that $r = 0.03$ and $t_0 = 0$.} 
\end{figure} 
Let the excited mode be the lowest frequency, $n = 1$, mode. It is possible to 
arrange for the observer to see net negative integrated energy for 
selected values of the parameters $r$, $v$ and $t_0$. For example, ${\cal E}$ 
is plotted in Fig.~\ref{fig:evt} as a function of $t_0$ for $r=0.03$ and $v=0.9$. 
The result can be either positive or negative. The cavity contains net positive 
energy, but with oscillatory pockets of negative energy density. An  observer 
who enters the cavity at certain times during the cycle will manage to see net 
negative energy, whereas one who enters at other times may see net positive 
energy. It is also of interest to look at ${\cal E}$ as a function of $v$ for 
fixed $r$ and $t_0$. This is illustrated in Fig.~\ref{fig:evs}, where 
$r=0.03$ and 
$t_0 = 0$. Note that for smaller values of $v$, ${\cal E} > 0$, whereas 
larger values of $v$ allow the observer to see net negative energy, 
${\cal E} < 0$. If the observer spends too long in the cavity, the energy 
oscillations cause the time-integrated energy to be positive, but a 
speedier observer can manage to catch net negative energy. Again it is 
important to emphasize that this net negative energy represents only the 
contribution of the quantum field in the cavity and not of the walls themselves. 
For any realistic cavity, it is overwhelmingly likely that the AWEC integral 
including the walls' rest mass energy will be positive.

\subsection{Strong and Weak Violations of the WEC in a Cavity} 
 
We have seen in Sec.~\ref{sec:explicit} that the scalar field in a single mode 
squeezed vacuum state can violate the WEC. For a travelling wave mode, it 
was shown that the violation is always weak for the massive field and 
always strong for the massless field. The cavity discussed in the previous 
subsection allows us to give an example where both strong and weak violations occur 
simultaneously in different regions of space. Again take a massless 
 scalar field in the 
cavity to be in a squeezed vacuum state for the mode given in Eq.~(\ref{eq:cavity_mode}). 
The energy density and pressure are equal and given by 
\begin{equation} 
T_{00} = T_{11} = A^2\,\omega^2\, \sinh r 
[\sinh r - \cos 2\omega x \, \cos 2\omega t \, \cosh r] \, , 
\end{equation} 
The flux is given by 
\begin{equation} 
T_{01} =  A^2\,\omega^2\, \sinh r\, \cosh r\, \sin 2\omega x \, \sin 2\omega t \,. 
\end{equation} 
 
\begin{figure} 
\includegraphics[scale=.5]{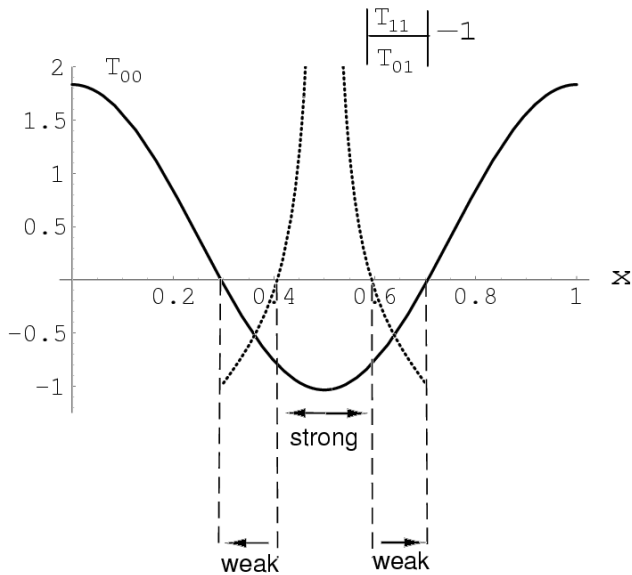} 
\caption{\label{fig:svw} The regions of strong and weak violation of the 
WEC are shown. The energy density and the parameter $|T_{11}/T_{01}|-1$ 
for the $n=1$ mode are plotted as functions of position in the cavity at 
time $t=3/8$. Here units in which $L=1$ are used, and we have set $A=1$ and 
$r=0.2$.} 
\end{figure} 
Let us suppose that we are at a point at which the WEC is violated, so 
$T_{00} < 0$, or $\cos 2\omega x \, \cos 2\omega t > \tanh r$. 
If $|T_{11}| > |T_{01}|$, we are in Case 1 of the Appendix, 
in which the necessary 
and sufficient condition for a strong violation is Eq.~(\ref{eq:pita}). 
However, when $T_{00}= T_{11} < 0$, this condition always holds if 
$|T_{11}| > |T_{01}|$. On the other hand, suppose that $|T_{11}| \leq |T_{01}|$. 
Then we are in Case 2.2 of the Appendix, and the necessary and sufficient 
condition for a strong violation is Eq.~(\ref{eq:nec_cond}). In summary, in 
the cavity all WEC violations are strong if $|T_{11}| \geq |T_{01}|$ 
and weak if $|T_{11}| < |T_{01}|$. It is possible to find both types 
of violation, as is illustrated in Fig.~\ref{fig:svw}. In this example, the 
WEC violation is strong in the middle of the $(-)$ energy region, and weak 
nearer to its edges. 
 
\section{\label{sec:summary}Summary and future directions} 
 
Let us summarize some of the results obtained in this paper, as well as 
some of the unanswered questions which this investigation has raised. 
We have given some explicit examples of spacetime averaged quantum inequalities 
in two-dimensional spacetime. However, the problem of finding similar results 
in four-dimensional spacetime is unsolved. We have used the AWEC and QIs 
to rule out or limit some particular model distributions of \n\  energy. 
In particular, the ``cap the cone'' argument given in 
Sec.~\ref{subsubsec:fcdistr:f:sep} shows that one cannot have a piece 
of \n\  energy separated from rigidly moving positive energy. We were able 
to give quantitative restrictions on other possible distributions. 
We also gave some explicit examples of allowed distributions. However, 
much more work needs to be done to narrow the gap between distributions 
which can be ruled out and those which are definitely allowed. As part 
of our investigation, we have introduced the distinction between strong 
and weak violations of the WEC, which is likely to prove useful in future 
work on this subject. We have also tested the limits of the AWEC and provided 
counterexamples to the AWEC along piecewise geodesics and to the difference AWEC 
for observers who pass through a cavity. These types of counterexamples are 
useful for understanding more clearly just which conditions can be used 
to constrain spatial distributions of \n\ energy. 
 
Future work in this area will involve a search for a more systematic ways to 
use information from worldline integrals to reconstruct or constrain 
spatial and spacetime distributions of \n\  energy. It will also involve the 
construction of additional explicit examples. It is especially interesting to 
see how far one can go in four spacetime dimensions toward having separated 
regions of \p\  and \n\  energy. 
 
\appendix* 
 
\section{\label{sec:a1}Strong and weak violations of the weak energy condition} 
 
If an observer measures negative energy at a point, 
must others measure it as negative too?  If all observers measure 
the energy at a point to be negative, 
we say that we have a strong violation of the weak energy condition at that point. 
If only some do, we say that we have a weak violation.  Suppose that an 
observer measures negative energy; i.e., $\T00<0$ in the observer's rest 
frame.  What are the conditions on the components of \T\mu\nu\ so that there 
is a strong violation of the weak energy condition?  We consider 
the question in two-dimensional flat spacetime. 
 
Under a Lorentz transformation, we have 
\begin{equation} 
T'_{00} = \gamma^2 (\T00 + 2 v \T01 + v^2 \T11), 
\end{equation} 
where $v$ and $\gamma$ are the usual boost and Lorentz 
factors, and $v$ obeys $-1<v<1$. 
Assuming that $\T00<0$, we want the necessary and sufficient condition that 
$T'_{00}<0$ as well, no matter what the value of $v$.  This occurs trivially, 
for instance, if both \T01 and \T11 are zero. We call this the trivial 
case. 
 
In nontrivial cases, the condition 
\begin{equation} 
|\T00| \geq \T11 + 2 |\T01| 
\label{eq:nec_cond} 
\end{equation} 
is necessary for $T'_{00}<0$, $\forall v$. 
In order to see this, suppose that the condition 
is violated. One of $\T01$ or $\T11$ cannot be zero, 
so we must have 
\begin{equation} 
-\T00 < \T11 + 2 |\T01|. 
\end{equation} 
This implies that for sufficiently small $\epsilon>0$, we have 
\begin{equation} 
-\T00 < (\T11-\epsilon/2) + 2 (|\T01| - \epsilon/4). 
\end{equation} 
Now, there must exist some $v_1>0$ such that 
\begin{equation} 
v^2\T11> \T11 - \epsilon/2, \quad\forall |v| > v_1, 
\end{equation} 
and some $v_2>0$ such that 
\begin{equation} 
|v||\T01|> |\T01| - \epsilon/4, \quad\forall |v| > v_2. 
\end{equation} 
Putting these together, we see that 
\begin{equation} 
\T00 + v^2\T11 + 2|v||\T01| > 0, \quad\forall v>\max(v_1,v_2). 
\end{equation} 
We choose $v$ to be \p\ or \n\ depending on whether 
$\T01$ is \p\ or \n. This gives us 
\begin{equation} 
T'_{00} = \T00 + v^2\T11 + 2v \T01 > 0, \text{\ for some\ } v. 
\end{equation} 
In other words, we do not have a strong violation of the weak 
energy condition. 
 
In order to discuss sufficient conditions, there are 
two cases we need to consider: 
\begin{enumerate} 
\item {$\T11<0$ and $|\T01|<|\T11|$} 
\item {Other} 
\end{enumerate} 
We look at the second case first. In this case, condition~\ref{eq:nec_cond} 
is sufficient as well.  To see this, we need to look at two subcases: 
 
\textit{Case 2.1, $\T11\geq 0$\/}: 
Since $|v|<1$, we have $\T11\geq v^2 \T11$ and 
$2|\T01|\geq 2|v||\T01|\geq 2v\T01$, with equality holding 
in each instance 
only if both sides are zero.  Since we are looking at nontrivial cases, 
at least one of \T11 and \T01 is non-zero. Then condition~\ref{eq:nec_cond} 
implies that 
\begin{equation} 
-\T00\geq \T11 + 2|\T01|>v^2 \T11 + 2v\T01, \quad\forall v. 
\end{equation} 
In other words 
\begin{equation} 
T'_{00} = \T00 + v^2\T11 + 2v \T01 < 0, \quad\forall v. 
\end{equation} 
 
\textit{Case 2.2, $\T11< 0$ and $|\T01|\geq|\T11|$\/}: We must have 
$\T01\ne0$ here, otherwise we get the trivial case. Define 
the function $f(v)$ by 
\begin{equation} 
f(v)\equiv \T00 + 2v\T10 + v^2\T11, \quad-1\leq v\leq 1. 
\end{equation} 
The graph of this function, under the imposed conditions, is a downward 
pointing parabola whose extremum $v_{\text{ext}}=-\T10/\T11$ lies 
outside the domain.  In order to have $f(v)<0$ for all $v$, 
we need to ensure that the higher of $f(1)$ and $f(-1)$ is nonpositive. 
When $\T01>0$, the higher of the two is $f(1)$ and when $\T01<0$, the 
higher point is $f(-1)$.  Now, condition~\ref{eq:nec_cond} reduces to 
$f(1)\leq 0$ when $\T01>0$ and to $f(-1)\leq 0$ when $\T01<0$, giving us 
precisely what we want. 
 
\textit{Case 1, $\T11<0$ and $|\T01|<|\T11|$\/}: 
Condition~\ref{eq:nec_cond} is still necessary here, but it 
is not sufficient. If, for example $\T00=-0.0001$, $\T11=-4$ and 
$\T01=1$, it is easy to check that the condition holds. Yet, for $v=1/200$ 
we get a positive value for $T'_{00}$. In order to derive the correct 
condition, consider the function $f(v)$ defined above. In this case, 
the extremum, $v_{\text{ext}}=-\T10/\T11$, lies inside the domain 
of the function and the condition to impose is $f(v_{\text{ext}})<0$. 
Therefore, the necessary and sufficient condition in this case is 
\begin{equation} 
\T00 < \frac{\T01^2}{\T11}. 
\label{eq:pita} 
\end{equation} 
This is a stronger condition than~\ref{eq:nec_cond}, in that it implies 
that condition but is not implied by it. 
 
The results of this appendix may be summarized in the following theorem:

\smallskip
\noindent {\bf Theorem:} Let $T_{\mu\nu}$ be the stress
energy tensor in a two                                                                                         -dimensional flat spacetime. Suppose that at
some point $P$ we have a negative energy density, i.e.,
$T_{00}<0$.  The conditions that
the energy density $T'_{00}$ in an arbitrary frame is also negative are
as follows:
\begin{enumerate}
\item {If $T_{01}=0$ and $T_{11}=0$, then $T'_{00}$ is automatically
negative.}
\item {If $\T11<0$ and $|\T01|<|\T11|$, then
\begin{equation}
|\T00| \geq \T11 + 2 |\T01|
\end{equation}
is necessary but not sufficient for $T'_{00}<0$. The condition
\begin{equation}
\T00 < \frac{\T01^2}{\T11}
\end{equation}
is necessary and sufficient for $T'_{00}<0$.}
\item {In all other cases
\begin{equation}
|\T00| \geq \T11 + 2 |\T01|
\end{equation}
is necessary and sufficient for $T'_{00}<0$.}
\end{enumerate}

These results may theoretically be applied to the four-dimensional case as well. 
Suppose that we have a violation of the weak energy condition in the 
rest frame of an observer.  Does an observer boosted in a spatial 
direction $\vec x$ also measure negative energy? 
We may rotate coordinates so that $\vec x$ points in the new x-direction 
(\T00 will be unaffected by the transformation), then apply the conditions 
of this section to the \T00, \T01 and \T11 components in the rotated 
coordinates. 
 
\begin{acknowledgments} 
We would like to thank Chris Fewster and Adam Helfer for useful 
discussions. This work was supported in part by the National 
Science Foundation under grants PHY-9800965 and PHY-9988464. 
AB and TR thank the Institute of Cosmology at Tufts University 
for its hospitality while this work was being done. AB also thanks the 
Faculty Research Awards Committee at Southampton College for 
partial financial support. 
\end{acknowledgments}

\end{document}